\documentstyle[12pt]{article}
\begin{document}
\def\be{\begin{equation}}
\def\ee{\end{equation}}
\def\ba{\begin{eqnarray}}
\def\ea{\end{eqnarray}}
\def\nn{\nonumber \\}
\def\pr{\prime}
\def\pp{\prime\prime}
\def\ub{\underbar}
\def\la{\label}
\def\re{\ref}
\def\i{{\rm i}}
\begin{titlepage}
\hfill   TUW--94--24 \\
\vfill
\begin{center}
{\Large Conserved Quasilocal Quantities}\\ \smallskip
{\Large and General Covariant Theories in Two Dimensions}
\vfill
\renewcommand{\baselinestretch}{1}
{W. Kummer\footnote{wkummer@tph.tuwien.ac.at} and P. Widerin}\footnote{ 
Address after Oct. 1, 1994: Institut f\"ur Theoretische Physik, 
ETH- Z\"urich, H\"onggerberg,\\           
CH--8093 Z\"urich, Switzerland}\\
Institut f\"ur Theoretische Physik \\
Technische Universit\"at Wien\\
Wiedener Hauptstr. 8-10\\
A-1040 Wien\\
Austria\\
\end{center}
\vfill
\begin{abstract}
General matterless--theories in 1+1 dimensions include dilaton gravity,
Yang--Mills theory as well as non--Einsteinian gravity with dynamical
torsion and higher power gravity, and even models of spherically symmetric
d = 4 General Relativity.  Their recent identification as special cases of
'Poisson--sigma--models' with simple general solution in an arbitrary
gauge, allows a comprehensive discussion of the relation between the known
absolutely conserved quantities in all those cases and Noether charges,
resp.  notions of quasilocal 'energy--momentum'.  In contrast to Noether
like quantities, quasilocal energy definitions require some sort of
'asymptotics' to allow an interpretation as a (gauge--independent)
observable.  Dilaton gravitation, although a little different in detail,
shares this property with the other cases.  We also present a simple 
generalization of the absolute conservation law for the case of interactions 
with matter of any type.
\end{abstract}
\vfill

Vienna, January 1995 \hfill 
\end{titlepage}

\renewcommand{\baselinestretch}{2}

\section{Introduction} 
The interest in two dimensional diffeomorphism invariant theories has many 
roots. Presumably the most basic one is the central role played by 
spherically symmetric models in d = 4 General Relativity (GR) as a 
consequence of Birkhoff's theorem. Two dimensional models with 'time' and 
'radius' possess an impressive history with promising recent developments 
\cite{ber}. On the other hand, the fact that the Einstein--Hilbert action of
pure 
gravity in 1+1 dimensions is trivial, also has spurred the development of 
models with additional nondynamical \cite{tei} and dynamical (dilaton and
tachyon) 
scalar fields \cite{bro}, besides higher powers of the curvature 
\cite{bro,ban}.
Especially the study of 2d--dilaton theories turned out to be an important 
spin--off from string theory, and has led to novel insights into properties 
of black holes \cite{bro,ban,man}. For a scalar field, coupled more generally
than a 
dilaton field, even more complicated singularity structures have been found 
\cite{lem} than in the ordinary dilaton--black hole \cite{bro}. \\
Actually such structures were known already before in another branch of 
completely integrable gravitational theories which modify 
Einstein--relativity in 1+1 dimensions by admitting nonvanishing dynamical 
torsion \cite{kata,katb}. Here the introduction of the light--cone (LC) gauge
led to 
the expression of the full solution in terms of elementary functions [9] and 
to an  understanding of quantum properties of such a theory in the 
topologically trivial \cite{kumb} and nontrivial \cite{sch} case. \\
Recently important progress has been made by the insight \cite{stra} that
\ub{all} theories listed above are but special cases of a
'Poisson--sigma--model' (PSM) with action 
\be 
    L = \int\limits_{M} (A_{B} \wedge d X^{B} + \frac{1}{2} P^{B C}(X) A_{B}
      \wedge A_{C})\;\; .  \la{1:1}
\ee
The zero forms $X^B$ are target space 'coordinates' with connection one 
forms $A_B$. $P$  expresses the (in general degenerate) Poisson--structure on 
the manifold $M$, it has to obey a Jacobi--type identity, generalizing the 
Yang--Mills case, where $P$ is linear in $X$ and proportional to the structure

constants. For the subclass of models describing 2d--covariant theories the 
$A_B$ are identified with the zweibein $e^a$, with the connection 
${\omega^a}_b  = {\epsilon^a}_b\,\omega$,  and may include possibly further
 Yang--Mills 
fields  $A_i$. Introducing the Minkowskian frame metric $\eta_{ab} = 
{\rm diag}\,(1,- 1)$,  target coordinates $X^A$ on the manifold 
will be denoted  as  $ \{X^a,X,X^i\}$ . In order to have a generic model 
including also another gauge field we shall consider in our present paper 
occasionally $X^i \to Y$ for a U(1)--connection $A_4 = A$. Of course,  a 
nonabelian Yang--Mills field could be included as well, even in our simple 
explicit model. \\
Then the (matterless) dilaton, torsion, $f(R)$--gravity theories including 
U(1) gauge fields \cite{wal} and even spherically symmetric gravity are
obtained 
as special cases \cite{strb} of an action of type (1), namely ($ \epsilon = 
\frac{1}{2} \varepsilon_{ab}e^a\wedge e^b $ )
\be
     L = \int\limits_{M} (X_{a} D e^{a} + X d \omega + Y d A - \epsilon V)\;.
     \la{1:2}
\ee
Appropriate fixing of $V = V(X^a X_a, X,Y)$ yields all the models listed 
above (and many more). It is possible in principle to write down the full 
solution for (\re{1:2}) in an arbitrary gauge (coordinate system). As seen
below, the solution has much of the shape of the LC gauge solution
\cite{kuma},  but in its present general form resembles more a generalization
of  H. Verlinde's solution to the (torsionless) dilaton case \cite{ver}. It
also  shows that the form of the solution given in \cite{ike} and \cite{sol}
for  nonvanishing torsion may be generalized further. \\
A crucial role for the integrability of theories (\re{1:1}) in the general case

play 'Casimir--functions'  $C_i(X^A)$ \cite{stra,strb} which on--shell become 
constants and thus (gauge--independent) 'observables',  also in the classical 
case \cite{int}. These constants $C_i$ together with other parameters in $P$
(or
$V$, e.g. the cosmological constant) determine the almost limitless variety of

Penrose diagrams, characterizing the singularities of such theories 
\cite{lem,katb,sol,klo}.  E.g. Schwarzschild or Reissner--Nordstr\"om black
holes  are just relatively simple members of that set. In a previous note
\cite{kumc}  we have related $C_1$ for the special theory \cite{kata} 
  to a global symmetry of the action which, of course, easily
generalizes  to the present framework. \\
Within the context of dilaton and dilaton--like models, conservation laws of 
total energy have been discussed repeatedly \cite{fro,sol,bil,geg}. Although
also  some Noether charge concepts of GR have been used \cite{naw}, the notion

'ADM'--mass \cite{arn} seems to cover various approaches, which, however,
always  basically  refer to variants of the Regge--Teitelboim argument 
\cite{reg} for canonically formulated theories in a finite space volume: Only
the choice of appropriate surface terms allows to drop the requirement of 
vanishing variation on those surfaces. Terms of this 
type are built in automatically into carefully developed concepts of
quasilocal energies (cf. e.g. \cite{yor}). \\
In 1+1 dimensions, of course, such a surface reduces to two points in a 
space--like direction. In addition, already for the relatively simple case of 
dilaton gravity, asymptotic flatness is achievable only in one
'space'--direction 
\cite{bro,bil} and the 'vacuum' still contains a dilaton field. This problem
even becomes more acute in the general framework of PSM-s with more general 
structure \cite{ber,tei}, where usually no direction leads into flat
asymptotics. Thus all GR approaches  based upon requirements of this type
become inapplicable. On the other hand, we now have a situation where the
conserved quantities are known. Thus, in our present work, we are able to turn
the problem around. We search a proper definition of a Noether 'charge' or 
quasilocal 'energy', if possible,  at a 
\ub{finite} 'distance' which in a gauge independent manner should reproduce 
the (in our case available and completely known) conservation laws. We also 
need not just compare a certain prescription with few examples, like the 
mass of the Schwarzschild black hole and its relatives in GR, but with 
conserved quantities for any singularity in an (almost limitless)  set of 
singularities which may be 'designed' at will by a suitable choice of $V$ in 
(\re{1:2}) (admittedly in $d = 2$ only). \\
In \underline{Section 2} we summarize the general solution, using for explicit
demonstration a model including $R^2$--gravity and dynamical torsion and
containing  also a U(1) Yang--Mills field. The reason for introducing the
latter
is to see how more than one 'Casimir--function' may yield one 'energy'. The 
nonpositivity of that energy is evident in all generic models. Among 
the different special cases the torsionless limit (including dilaton gravity)
will be of special interest. We then discuss the general Killing field which
belongs to a metric formed by the solution in an arbitrary gauge. \\
The analysis of different ways to determine the energy by means of surface 
terms in \underline{Section 3} starts with  the concept of 
'energy--momentum' by analogy with the case involving matter in d = 4 for 
nonvanishing torsion \cite{ike,wal}.  Then
 the 'classical' Noether--procedure follows which we combine
 with the ADM--setting for space and time. We also apply the general 
 formulation of Wald \cite{wal} which covers all covariant theories. In each 
 case dilaton gravity shows differences in detail, but not in principle. \\
 A common difficulty of all these concepts is the interpretation of the 
 'mass--shell', i.e. the manner the equations of motion have to be used. \\
A second basic problem arises, if one tries to implement a quasilocal energy 
concept in a gauge--independent way, without having an asymptotically flat 
space at disposal. These problems can  be addressed directly in terms of the
Regge--Teitelboim argument.  One approach follows the trick of recombining
constraints \cite{geg,fis}.  
The identical result for the surface term can be obtained, however,
also directly.  Moreover, it is now possible to apply a simple consistent
prescription which solves the 'mass--shell' problem in a very
straightforward manner.  Still, the quasilocal energy related to that
surface term --- as in the previous approaches --- only yields the
conserved quantity in some 'asymptotic' sense.  \\
It should be noted that all our results refer to the classical theory. 
Nevertheless, precisely the Casimir functions in the matterless case become
the discrete quantum variables living on a compactified space $ S^1 $
\cite{sch,stra,strb}.  Sparse remarks on that and on the changes in the case
of
interactions with matter are included in the final outlook
(\underline{Section 4}).

\section{PSM--Gravitation}
\subsection{The General Model}

With $ V $ in (\re{1:2}) depending linearly on $ X^{a} X_{a} $
\be
   V = \frac{\alpha}{2} X^{a} X_{a} + v(X,Y)\;\; ,
     \la{2:1}
\ee
the equations of motion from (\re{1:2}) in a LC basis of the frame metric 
( $ \varepsilon_{+-} = -1$ , ${ X^{\pm} = (X^0 \pm X^1)/\sqrt{2}}$,
$ \eta_{+-} = \eta_{-+} = 1 $) are
\ba
     d X^{\pm} \pm \omega X^{\pm} &=& \pm e^{\pm} V \nonumber\\
     d X + X^{-} e^{+} - X^{+} e^{-} &=& 0 \la{2:2}\\
     d Y &=& 0  \nonumber
\ea
and
\ba
  d e^{\pm} \pm \omega \wedge e^{\pm} = - \alpha e^{+} \wedge e^{-} X^{\pm}
\nn
  d \omega = - e^{+} \wedge e^{-} \frac{\partial v}{\partial X}  \la{2:3}\\
  d A = - e^{+} \wedge e^{-}  \frac{\partial v}{\partial Y}\;\; .\nonumber
\ea
Multiplying the first pair of equations in (\re{2:2}) with $X^-$ and $X^+$, 
respectively, the second one with V and adding yields 
\be
   d(X^{+}X^{-}) + V d X = 0, \la{2:4}
\ee
producing an absolutely conserved quantity ($d\, C = 0$)
\ba
    C_{1}= C =  X^{+}X^{-} e^{\alpha X} + w(X,Y) 
   \la{2:5}\\
   w(X) = \int\limits_{X_{0}}^{X} v(y,Y) e^{\alpha y} dy\;\; . \la{2:6}
\ea
Clearly the lower limit $X_0 = const.$ must be determined appropriately so that

(inside a certain patch) the integral exists for a certain range of the 
(curvature) $X$. Eq. (\re{2:5}) generalizes \cite{stra,strb} the previously
known 
\cite{kata,kuma} analogous quantity for 2d gravity with dynamical
torsion. However,  the limit $\alpha \to 0$ immediately also yields the
conservation law for torsionless cases (F(R)--gravity, dilaton gravity
\cite{ver} etc.). With our additional U(1)--field, (\re{2:5}) depends on the
'field--strength'  which according to (\re{2:2}) is the second conserved
quantity  $ C_2 = Y$. \\
Setting one LC component of the torsion, e.g. $X^+$ identically zero, the 
first equation (\re{2:3}) (at $e^+ \neq 0$) may yield a constant curvature 
depending on $Y$, if $v(X,Y) = 0$ has a solution at all. Such 'de--Sitter' 
solutions are related to $ C = 0$ (within an appropriate convention for the 
integration constant in (\re{2:6})). Representing discrete points in phase 
space they are notorious  especially in the quantum case \cite{sch} but, 
fortunately, need not be considered in our present context. If $X^+ \neq 0$ 
the solution of (\re{2:2}) and (\re{2:3}) becomes 
\ba
     e^{+}  & = & X^{+} e^{\alpha X} d f\nn
     e^{-}  & = & \frac{d X}{X^{+}} + X^{-} e^{\alpha X} d f \nn
     \omega & = & - \frac{d X^{+}}{X^{+}} + V e^{\alpha X} d f 
     \la{2:7}\\
     A      & = & d g  + \frac{\partial w}{\partial Y} d f\nonumber
\ea
in terms of arbitrary functions $f, X, X^+$. One simply solves one of 
the first eqs. (\re{2:2}) for $\omega$, the second eq. for, say, $e^-$ and  
inserts into (\re{2:3}), using (\re{2:4}) etc. This means that (\re{2:7}) is
understood  with $X^-$ in $e^-$ and in $V$ to be reexpressed in terms of $C$
by (\re{2:5}).  \\
Of course, for $X^- \neq 0$ the analogous solution exists with the roles of 
$ X^+ \leftrightarrow X^-$ exchanged. This is important for patching 
together general solutions \cite{stra,strb}.\\
The first terms in the first three eqs. for $ e^{\pm} \rightarrow \delta
e^{\pm}, \omega \rightarrow \delta \omega, d f \rightarrow \delta \gamma  $
are the on--shell extension of a global nonlinear symmetry of (\re{1:2}),
\cite{kumc}. It is related to the conservation $\partial_\mu\, J^\mu_\nu = 0$ 
of a Noether current $ J^\mu_\nu = C (X^aY_a, X,Y)\,\delta^\mu_\nu$ 
because under such a transformation the Lagrangian density in (2) changes by 
a total derivative only.  
Mathematically (\re{2:7}) coincides with the solution in the LC--gauge
\cite{kuma,he}  where the curvature $X$ is gauge--fixed to 
be linear in 'time'. But (\re{2:7}) has the big advantage that it is valid in
an arbitrary gauge, whereas solutions obtained in the literature to such 
theories had to rely on special gauges and sometimes on sophisticated  
mathematical methods to solve the respective equations (cf. e.g. 
\cite{ban,man,lem,kata,geg}). The line element from (\re{2:7}) generally reads

\be
   (ds)^{2} = 2 e^{\alpha X} d f \otimes (d X + X^{+} X^{-} e^{\alpha X} d 
   f)\;\; ,\la{2:8}
\ee    
with $X^+X^-$ to be expressed by (\re{2:5}) for fixed $C$. For the case with 
torsion our generic model with a U(1) field may be chosen as
\be
   V = \alpha X^{+}X^{-} + \frac{\rho}{2} X^{2} + \sigma X Y + \frac{\tau}{2}
      Y^{2} - \Lambda\;\; . \la{2:9}
\ee
This $V$ allows to produce $C$ by a simple integral according to (\re{2:5}).
Integrating out $X$ and $X^\pm$ in (\re{1:2}) for $\sigma = \tau = 0$ leads to
the model quadratic in curvature and torsion of \cite{kata,kuma} which, in
four dimensions, together with the Einstein--Hilbert term has been known as
the 'Poincare--gauge theory' for some time \cite{ike,he}. It only contains
second 
derivatives in the field equations for the variables $e^a$ and $\omega$. 
However, higher derivative theories are to be treated with equal ease, when 
polynomials of higher degree in $X$ and $X^+X^-$ are admitted in (\re{2:1}). Of

course, $V$ could even be a nonpolynomial function. This 
would only make the integration harder which leads to (\re{2:5}). As we shall
recall 
shortly below, the zeros of (\re{2:5}) determine the singularity structure of
the theory. Thus one could design such a structure by prescribing  
$ C (X^{A})$ . 
{}From (\re{2:5}) the corresponding $V$ can be read off by differentiation, and
the 
action for that structure follows immediately. \\
Among the models with vanishing torsion ($\alpha = 0$ in (\re{2:1})), the 
Jackiw--Teitelboim model obtains for $v = \Lambda X$. Witten's black hole 
\cite{bro} represents a special case of a class of more general torsionless 
theories involving the curvature scalar R and one additional scalar field 
\cite{ban,geg} in a Lagrangian of the type 
\be
    {\cal L} = \sqrt{-g}[ \partial_{\alpha} \varphi \partial_{\beta} \varphi
               g^ {\alpha\beta} + A(\varphi) + R B(\varphi)] \la{2:10}
\ee
with arbitrary functions A and B. Matterless dilaton--gravity \cite{bro} is
the special case $\varphi^2 = 4B = A/\lambda^2 = 4\,e^{-2\Phi}$
\be
    {\cal L}_{dil} = \sqrt{-g} e^{- 2 \Phi} [4 \partial_{\alpha} \Phi
      \partial_{\beta}  \Phi g^{\alpha \beta} + 4 \lambda ^{2} + R]\;\;
.\la{2:11}
\ee
Using the conformal identity for $\tilde{g}_{\alpha\beta} = e^{-2\phi} 
g_{\alpha\beta}$ (or $\tilde {e}^a = e^{-\phi}\, e^a$)
\be 
     \sqrt{-\tilde{g}} \tilde{R} = \sqrt{-g} R + 2 \partial_{\alpha}
     (\sqrt{-g} g^{\alpha \beta} \partial_{\beta}\phi)\;\; .\la{2:12}
\ee
Eq. (\re{2:12}) allows the elimination of the kinetic term for $\varphi$ in 
(\re{2:10}) \cite{ver}. The resulting action may be written readily  in the
first order  form (\re{1:2}) for $\tilde e^a = e^{-\Phi} e^a, V = +4 \lambda^2,
X = 2 e^{-2\Phi}, Y =0$. Going back from (\re{2:7}) for $\alpha = 0$ as 
$ e^a = \tilde e^a / \sqrt{X/2}$ 
\cite{stra},  simply leads to
\ba
      e^{+} & = & X^{+} e^{\Phi} d f\nn
      e^{-} & = & \frac{1}{X^{+}}[-4 d \Phi e^{-\Phi} + d f (C\, e^{ \Phi}
                   - 8 \lambda^{2} e^{-\Phi})] \la{2:13}\\
      \omega & = & - \frac{d X^{+}}{X^{+}} + 4 \lambda^{2} d f\;\; . \nonumber
\ea
Here $\Phi, f$ and $X^+$ are arbitrary functions. E.g. the Kruskal form for 
the metric $(ds)^2 = 2e^+ \otimes e^-$ a dilaton black hole follows from the 
gauge--fixation $(X^+X^- = uv)$
\begin{eqnarray*}
     8 \lambda^2 e^{-2 \Phi} & = &  C - uv \\
     4 \lambda^{2} f & = &  \ln (u)\;\; .
\end{eqnarray*}
The mass of the dilaton black hole is related to $C$ by $ C = 8 \lambda M $.
$X^+ \neq 0$ being still arbitrary, it may be used to gauge $\omega = 0$ 
which shows that the connection $\omega$ in (\re{2:13}) really has nothing to
do with a curvature belonging to the metric derived from that equation. \\
Now precisely the same procedure may be applied to the generalized theories of

type (\re{2:10}). Here $\varphi$ can be eliminated \cite{ban,geg} using
(\re{2:12}) 
\be
  \tilde{{\cal L}}= \sqrt{-{\tilde g}}[A(\varphi)/F(\varphi) + \tilde{R}
B(\varphi)]
                                          \la{2:14}
\ee
with 
\begin{eqnarray}
     g_{\alpha \beta} & = & \tilde g_{\alpha \beta} / F(\varphi) \nonumber\\
     \ln\, F(\varphi) & = &  \int\limits^{\varphi} dy/(\frac{d B}{d 
     y}) \;\; .\la{2:15}
\end{eqnarray}
The corresponding first order action (\re{1:2}) becomes
\begin{eqnarray*} 
   \tilde{L} = \int (X_{a} D \tilde{e}^a + 2 B d \omega - \tilde{\epsilon} 
   A/F)\;\; .	\nonumber
\end{eqnarray*}
In (\re{1:2}) for $\alpha = 0$ we have as a consequence

\parbox{12.0cm}{
\begin{eqnarray*}
    X & = & 2 B \\
    V = v & = & + \frac{A(B^{-1}(X/2))}{F(B^{-1}(X/2))} \;\; ,
\end{eqnarray*}}\hfill\parbox{1.0cm}{\begin{eqnarray}\la{2:16}
\end{eqnarray}}\\
and the conserved quantity for any theory of type (\re{2:10}) is (\re{2:5})
with 
$\alpha = 0$ and
\be
    w(X) = + \int\limits_{B^{-1}(X_{0}/2)}^{B^{-1}(X/2)} \frac{A(y) d 
    y}{F\,(y)}\;\; .\la{2:17}
\ee
The line--element in terms of coordinates $(f,X) = Y^\alpha$  reads for any
such theory and in any gauge
\ba
        (d s)^{2} = F^{-1} 2 d f \otimes [dX + df ( C - w(X))]\;\; .\la{2:18}
\ea

Of course, in each application to a particular model a careful analysis of 
the range of validity of the mathematical manipulations is required in order 
to determine a patch, where those steps are justified: allowed ranges for 
transformations of fields, inversions of functions like $B^{-1}$, 
integrability of $F$, admissible gauges for $f$ and $X$ etc. It is precisely 
for this reason that the very comprehensive framework reviewed here
\cite{stra} does not invalidate approaches in special gauges for special models

\cite{ban,sol,geg} which allow a careful analysis of these points. \\
Another example is the action for the Schwarzschild black hole in 4d GR. 
$v(X) = - 1/(2 X^2)$ in (\re{1:2}) is found to yield the correct line--element

\cite{stra}. \\
In the literature also theories with vanishing curvature and dynamical 
torsion have been considered ('teleparallelism' theories \cite{ike,he}). E.g.
a
pure $T^2$--action is the limit $\rho = \sigma = \tau = 0 (\alpha \neq 0)$ in 
(\re{2:8}).

\subsection{Killing Vector and Singularities}

In our very general class of models the Killing vector can be found  
without fixing the gauge (coordinate--system). 
Using (\re{2:7}) we rewrite the line element  (\re{2:8}) in a 
theory (\re{1:2}) as 
\be
     (d s)^{2} = d f \otimes [ 2 e^{\alpha X}d X + l d f]\la{2:19}
\ee
where
\be 
    l = 2 X^{+} X^{-} e^{2 \alpha X} = 2 e^{\alpha X}(C - w(X,Y))\;\;
.\la{2:20}
\ee
In terms of the variables $Y^\alpha = (f,X)$, resp. $\partial / \partial 
Y^\alpha , k^\alpha $  is the Killing vector with norm (\re{2:20}) 

\ba 
   k^\alpha &=& (1,0)\nonumber\\
   k^2 &=& k^{\alpha} k^{\beta} g_{\alpha \beta} = l\;\; .\la{2:21}
\ea
Hence $ l > 0$ in (\re{2:20}) determines a timelike Killing field. It is 
instructive to consider a translation in the ('on--shell') Killing direction 
$ - \delta \beta k^{\mu}$

\ba
       \delta e^{a}_{\nu} &=& \delta \beta k^{\mu} e_{\nu,\mu}^{a} + (\delta
              \beta k^{\mu})_{,\nu} e_{\mu}^a =\nn
          &=& (\delta \beta)_{,\nu} k^{\mu} e_{\mu}^a \;\; ,\la{2:22}
\ea
and the analogous equation for $e_\nu^a \to \omega_\nu$, inserting back 
$k^\mu e^a_\mu$, resp. $k^\mu \omega_\mu$ from the solution (\re{2:7}):
\ba
       \delta e^{a}_{\nu} &=& (\delta  \beta)_{,\nu} e^{\alpha X} X^a \nn
       \delta \omega_{\nu}  &=& (\delta \beta)_{,\nu} V  e^{\alpha X} 
       \la{2:23}
\ea
Comparing (\re{2:23}) to the global symmetry  in such theories \cite{kumc}
\ba
    \delta_{k} e^{a}_{\nu} &=& \delta  \gamma_{\nu} e^{\alpha X} X^a \nn
    \delta_{k} \omega_{\nu}  &=& \delta \gamma_{\nu} V  e^{\alpha X} 
    \la{2:24}\\
     \delta_{k} X^a &=& \delta_{k} X = 0\nonumber 
\ea
for a {\sl global} variation $\delta\gamma_\mu =
\epsilon_{\mu\nu}\delta\gamma^\nu$, we find complete agreement, as long as the
components $X^a = (X^+,X^-)$  in (\re{2:7}) are taken to be independent
functions
(and  not related by $C$  according to (\re{2:7})). For the discussion of the
singularity  structure of (\re{2:7}) a (partial) gauge fixing is useful. If
$ l > 0 $ in (\re{2:19}) we choose coordinates time $(t)$ and space $(r)$ in $
f =
f(t,r), X = X(r)$ with $\dot f = T(t)$ and
\be
  X' e^{\alpha X} + f' l(X) = 0\;\; ,\la{2:25}
\ee
where $f' = \partial f / \partial r \;\; ,$ \quad $ \dot f = \partial f /
\partial
t$. Introducing 
\be
   K(z) = - \int\limits_{z_o}^{z} d y e^{\alpha y} l ^{-1} (y)\;\; ,\la{2:26}
\ee
(\re{2:25}) implies
\be 
   f = \int\limits^{t} T(t') d t'+ K(X(r))\;\; .\la{2:27}
\ee
In such a gauge $g_{t r}$, the off--diagonal part of the metric 
vanishes, so that `space` and `time` are  separated.  
In order to  avoid zeros in the norm of the Killing--vector field $k$ it is
obvious to  restrict $z$ and $z_0$ to a suitable interval of $y = X(r)$
where $k$  exists. The remaining elements of $g_{\alpha \beta}$  are:
\ba
  g_{t t} &=& \dot{f}^{2} l\nn
  g_{r r} &=& - (f')^2 l\la{2:28}
\ea
Requiring a 'Schwarzschild'--form of the metric, i.e. $\det g = -1$, 
eliminates the arbitrary functions $T(t)$ and $X(r)$ altogether,
\ba
   T &=& 1\nn
  \alpha\, X &=&  \ln\,(\alpha r)  \;    (\alpha \ne 0)\la{2:29}\\
   X &=& r \quad (\alpha = 0)\;\; ,\nonumber
\ea
dropping a multiplicative constant $a$ in $f$, and $1/a$ together with $r$,
and two further constants for the zero points of $t$ and $r$. Now 
\be
  g_{t t} = - g_{r r}^{-1} = l(X(r))\la{2:30}
\ee
follows with $ X(r) $ from (\re{2:27}). Especially (\re{2:30}) clarifies the
remark  above, how an action may be reconstructed for a given singularity in
the  metric, proceeding backwards through (\re{2:20}) to (\re{1:2}). \\
We note that for a (generalized) dilaton theory, besides $\alpha = 0$, 
because of the additional factor $1/F$ in (\re{2:18}) there is a 
corresponding change to $l$ in (\re{2:28}) etc. Thus the singularity 
structure is determined by $l / F$. \\
The Katanaev--Volovich model with (\re{2:7}) at $\sigma = \tau = 0$ is 
sufficiently general to show the intrinsic singularity structure by an 
analysis of completeness of geodesics. $C^2$ global completeness was first 
shown in \cite{katb} within the conformal gauge. The more suitable LC--gauge
allows the extension to $C^\infty$ completeness and a discussion of possible 
compactifications \cite{sol,klo}. In that model altogether 11 types of Penrose

diagrams appear (G1, $\ldots$  G11 in the classification of \cite{katb}). 
Some show similarities to Schwarzschild and to Reissner--Nordstr\"om types, 
but there are many more.  In the more complicated cases they are obtained by
the  possibility to successively gluing together LC--solutions (I) and (II)
with  'complementary' gauges 
(LCI: $X^+ = 1, X = t, \omega_t = 0, X^- = X^-(C, X = t)$; LCII:$X^- = 1, X =
r,
\omega_r = 0, X^+ = X^+ (C,X=r)$) 
in appropriate lozenges. The diffeomorphisms for doing that is essentially
(\re{2:26})  again. For further details we refer to the relevant 
work \cite{lem,katb,sol,klo}.\\
 It is sufficient for our present purposes to note that for 
all types of singularities (including also e.g. naked ones) there are 
space--like directions allowing the study of surfaces (points) between such 
singularities at finite (incomplete case) or infinite (complete) distances.
Also a
second point is obvious from this section: In all two dimensional theories the
conserved  quantity (-ies) never has (have) a well-defined sign. Thus any hope
to find a positive `energy` must be in vain. Therefore, also adding matter to
the theories (\re{1:2}) is not likely to improve this situation.

\newpage

\section{Conserved Quantities and Surface Terms}

\subsection{Energy--Momentum}

 An intrinsic Palatini--type formulation of a covariant theory like (\re{1:2})

 precludes the immediate application of those well--tested concepts from GR 
 which are based upon $g_{\alpha\beta}$ as dynamical variables. The 
 (off--shell) expression $(e = det e_{\mu}^a)$
\be
    {\cal T} ^{\mu \nu}:= e^{a \nu} {\cal T}^{\mu}_a = \frac{e^{a \nu}}{e}
    \frac{\delta  L}{\delta e_{\mu}^a} \la{3:1}
\ee
clearly could be considered the analogue of the `energy  momentum tensor`, 
but the symmetry $\mu \leftrightarrow \nu$, needed e.g. to prove 
$({\cal T}^{\mu\nu}k_\nu)_{;\mu} = 0$ in the presence of a Killing--vector
$k_\mu$ cannot be true if applied naively \cite{sol}.  One
 more sophisticated approach to a conserved 'energy', also related to an
argument involving the Killing--vector, has been suggested in connection with
theories involving torsion in $d=4$ (cf. the second ref. \cite{ike},\cite{he}).

For that purpose let us
add to (\re{1:2}) an additional piece $L^{(m)}$ containing, say, scalar
fields. Then still $\delta L^{(m)}/ \delta \omega_{\mu} = 0$, and the only
change in the field equations occurs on the r.h.s. of the first eq. (\re{2:2})
where ${\cal T}^{(m)}_{a}$ (analogous to ${\cal T}^{\mu}_{a}$ in 
(\re{3:3})) may be written as 1-form ($a=+,-$): 
\be
  d X_a - \varepsilon_{b a} X^b \omega + e^b \varepsilon_{b a} V =
  {\cal T}^{(m)}_a \la{3:2}
\ee
The rest of (\re{2:2}) remains unchanged.  By analogy to GR we may assume
that matter is concentrated somewhere in space so that at the point at
which one considers the energy, the matterless equations (\re{2:2}) and
(\re{2:3}) hold 'asymptotically'.  Although this background for the effect
of matter is by no means flat, it possesses (within certain regions) a
time--like ($\ell > 0$) Killing vector (\re{2:21}) which does not depend on
space ($X$), despite the possible singularity structure of such models, as
reviewed at the end of Section 2.  Then it is natural to interpret 'energy'
--- for the matter part on the r.h.s.  of (\re{3:1}) but then also for
the l.h.s.  --- as the projection of ${\cal T}^{(m)}_a$ onto $k^a$, the 
matter--less, and hence also 'asymptotical' 
Killing field $k^a$ measured in terms of the inverse zweibein
($\hat{e}^{\alpha}_a e^b_{\alpha} = \delta^b_a$)

$$ 
   k^{\alpha} = \left( \begin{array}{c} 1\\0 \end{array}\right) = k^a
   \hat{e}^{\alpha}_a .           
$$
Using the $df$  components in (\re{2:8}) 
$$
   k^a = e^a_{\alpha} k^{\alpha} = e^{\alpha X} X^a     
$$
follows. Thus with (\re{3:2}) we have
\ba
  {\cal T}^{(m)} := k^a  \cdot  {\cal T}^{(m)}_a &=& (X^a dX_a + e^b
  \varepsilon_{b a} X^a V)e^{\alpha X}\nn
  &=&(d(X^+ X^-) + V dX )e^{\alpha X} = d\, C \la{3:3}
\ea
where the definition of $C$ in (\re{2:5}) has been inserted.  However, here
$C$ still must be considered to be a function of $X^+X^-, X\; {\rm and}\;
Y$ as in (\re{2:5}) and {\sl not} a constant, although this is implied by
the full field equations which contain (\re{2:4})! Disregarding this
problem for the time being (for the matter--less case) which we shall encounter

again below,from (\re{3:3}) for the matter part $d {\cal T}^{(m)} = 0$ or
\be 
     {\cal T}^{(m)} = - d\, J \la{3:4}
\ee
follows. Thus a conversation law holds, 

\be
d\, (C + J) = 0 \;\; ,
\la{3:4a}
\ee
involving the matter--less ('background') Killing projection in $J$.
The r.h.s. of (\re{3:3}) for the matter--less case implies -- what we really
already know from (\re{2:5}) -- that the 'Noether charge' $C = const $. 
Although we have
obtained a 'gauge--independent' definition of energy, our argument is not
conclusive for vanishing  matter (cf. e.g. also \cite{sol}), because
the r.h.s. of (\re{3:3}) with $C$ vanishes identically.\\
According to Sect.2,  dilaton  gravity and its generalizations differ from the

other  models by a 
redefinition of the zweibein by a factor involving the dilaton field 
according to (15) or a factor $\sqrt{F}$ in (\re{2:10}) with (\re{2:14}). 
This, however, does not change the preceding argument: 
${\cal T}^{(m)}_a$ in (\re{3:2}) just acquires  a further factor 
$\sqrt{F}$ upon variation 
with respect to the 'true' zweibeins $e^a$, instead of $\tilde{e}^a = 
\sqrt{F}\,e^a$, but the projection of the Killing--vector onto 
$e^a$ contains a compensating $1 / \sqrt{F}$. \\
Although starting from the consideration of a local object like ${\cal 
T}^{(m)}_a$, we thus arrive generically at surface--related quantities like 
$C$ and $J$. For 1 + 1 dimensional theories the  space--like 
boundary reduces to two points in space attached to two time--like curves. 
For any application to a singularity in 'space', enclosed by such a surface, 
those two space points also sit on a spacelike  curve which may even be
separated 
completely into two disconnected pieces  precisely by the singularity (and/or
a horizon) under consideration. 
Therefore, a 'surface integral ' consisting of the difference of the values 
at those points seems to be of little significance. It is rather the value 
at each point that is important. As we have seen already for the matter part 
in the preceding example, a 'superpotential' $J$  automatically leads to such 
surface terms as $Q^0 = \int dr\, \epsilon^{01}\, \partial_r J$ receives 
contributions from the boundary only. \\
Another important point is the question of the background \cite{yor}.
 In d = 4 GR for the black hole
the latter is represented naturally by  flat asymptotic space which also
determines a corresponding time--like Killing vector.  In the case with 
matter in d = 2 the corresponding background turned out to have a 
generically very complicated {\sl intrinsic} singularity structure with no 
direct relation to the matter contained in it, the Killing vector being 
related just to that (matter--less) background. This clearly points towards a 
possible basic weakness of any analysis based upon 2d covariant singular 
models, when a comparison with GR in d = 4 is intended. Still, within the d 
= 2 matterless models and their conserved quantity $C$ the freedom 
persists to subtract yet another  background according to some guiding
principle. 
We shall return to that point in the next subsection.

\subsection{Noether--Currents}

Among the pseudo--tensor approaches we take in (\re{1:2}) the simplest one with

text--book Noether currents for the (global) SO(1,1) Lorentz invariance 
$\delta a^\pm = \pm\delta\gamma a^\pm$ and (global) translation invariance 
$\delta x^\mu = \delta a^\mu = const$.  In coordinates $x^\mu = (t,r)$ the 
Lorentz--current becomes 
\be 
  J^{\mu} = \varepsilon^{\mu \nu} \partial_{\nu} X
                                         \la{3:5}
\ee
and the 'energy momentum tensor' 
\be
   {\cal T}_{\nu}^{\mu}= \varepsilon^{\mu \alpha}\partial_{\alpha} K_{\nu}
                                         \la{3:6}
\ee
\be
   K_{\nu} = X^+ e^-_{\nu} + X^- e^+_{\nu} + X \omega_{\nu} + Y A_{\nu}\;\; .
                                         \la{3:7}
\ee
Both expressions follow from the use of the field equations (the second eq. 
(\re{2:3}) 
in (\re{3:6}) and the first and last eq.(\re{2:3}) in (\re{3:7})) in the 
standard definitions of such currents, and both turn out to be expressible 
in terms of superpotentials. The corresponding
densities  $J^0 = \partial_1 X, {\cal T}^0_0 = \partial_1K_0$, upon
integration yield surface terms $X$, resp. $K_0$. Without any physically
motivated  choice of coordinates in (\re{3:6}) and (\re{3:7}) no relation to
an
'energy ' concept can be expected. Their possible significance may be checked
in  specific gauges, e.g. in LC gauge $e_0^+ = \omega_0 = A_0 = 0, e_0^- = 1$:
In that case the solution in (\re{2:7}) implies $X^+ = A(r), X = A(r)t +
B(r)$. However, a residual gauge--fixing \cite{kuma} allows to set $A = 1, B =
0$ for  which $\partial_1X = \partial_1 K_0 = 0$ follows {\it identically}. \\
In a general gauge with $\nu = 0 = $ 'time', (\re{3:7}) with (\re{2:7}) is
easily 
evaluated with the typical $V$ of (\re{2:8}) (e.g. for $\alpha \ne 0$):
\ba
      K_0 &=& \dot X - X \frac{\dot X^+}{X^+} + \dot g Y + \dot f \tilde E 
      \la{3:7a}\\
      \tilde E &=& 2 C (1 + \frac{\alpha}{2}X) + \frac{e^{\alpha
                  X}}{\alpha}(- \frac{2 \rho}{\alpha^2} + \frac{\rho X +
                  \sigma Y}{\alpha}+ 2 \Lambda)\la{3:8}
\ea
In an ADM--setting 
\be
   (ds)^2 = N^2 d^2 t - h (dr + N_1 dt)^2\la{3:9}
\ee
the situation for the surface term (\re{3:7}) is less trivial than in the LC 
gauge. With mutually orthogonal time--like and space--like vectors 
\ba
    n^{\pm \mu} n^{\pm}_{\mu} = \lambda_{\pm} = \pm 1\nn
    n^{+ \mu} n^-_{\mu} =0\la{3:10}
\ea
on 'hypersurfaces' (curves) with internal metric 
\be
   \gamma^{\pm}_{\mu \nu} = g_{\mu \nu} - \lambda_{\pm}n^{\pm}_{\mu}
   n^{\pm}_{\nu }  \la{3:11}
\ee
orthogonal to $n^{+\mu}$, resp. $n^{-\mu}$, a finite region in 1 + 1
space--time may be 
surrounded (as in the case d = 4 \cite{yor}). From (\re{3:9}) we obtain
\ba
  n^{- \mu} = \frac{1}{\sqrt{h}}\delta^{\mu}_1 &,& n^-_{\mu} =
  -\sqrt{h}\left( \begin{array}{c} N_1\\1 \end{array}\right)\nn
  n^+_{ \mu} = N\delta_{\mu}^0 &,& n^{+\mu} =
  \frac{1}{N} \left( \begin{array}{c} 1\\-N_1 \end{array}\right)\;\;
.\la{3:12}
\ea
The boundary with normal in the space--like direction simply becomes
\be
   \gamma^-_{\mu \nu} = \left( \begin{array}{cc} N^2 & 0\\0 & 0
    \end{array}\right)  \la{3:13}
\ee
For our solution (\re{2:7}) we conclude from (\re{2:8}) with (\re{2:20}) 
\ba
     \sqrt{h} N &=& e^{\alpha X}(\dot f X' - f' \dot X)\nn
     - h &=& 2 f' e^{\alpha X}(X' + f' \frac{l}{2} e^{-\alpha 
     X}) \la{3:14}\\
     - h N_1 &=& e^{\alpha X}(\dot f X' + f' \dot X + \dot f f' l
     e^{-\alpha X})\;\; .\nonumber
\ea
Again dots and primes  refer to derivatives with respect to $t$ and 
$r$. Here both $f$ and $X$ are functions of the (arbitrary) coordinates $t$ 
and $r$. Without restricting the generality of (\re{3:14}) we may decide to 
measure 'space' in terms of (curvature) $X$,  i.e. $\dot X = 0$. Still, 
(\re{3:14}) expresses altogether three functions $N, h$ and $N_1$ in terms 
of $X(r)$ and $f(t,r)$. For the proper definition of a quasilocal quantity 
on the surface, $\delta / \delta N$ will be needed at fixed $h$ and 
$N_1$. As there must be thus a relation between the three ADM--functions in 
our model it is cumbersome to perform that derivative in the general case. 
We, therefore, fix the gauge {\sl in part} to simplify matters, but in 
such a way that gauge(--in)dependence of results may still be checked. The 
most natural gauge fixing is $N_1 = 0$, i.e. (\re{2:25}),  as in section 2.2,

yielding (29).   The differentiation with 
respect to $N$ at fixed $h\, (r)$ implies differentiation at constant $r$.
By analogy with the  quasi--local definition of energy in ref. \cite{yor}
in that  case the projection 

\be
E = \frac{2 n_0^+ n_0^-}{\sqrt{\gamma}}\;\frac{\delta 
K_0}{\delta\gamma^-_{00}}\;\; ,\la{3:14a}
\ee
seen by a physical (geodesic) observer \cite{mas}
should be related to the 'energy' at fixed $r$. Note that $\gamma$ is the 
determinant of the submatrix in (\re{3:13}), i.e. $N^2$ again. We are, of
course, 
aware of the fact that this approach only finds its justification using the 
{\sl full} quasilocal approach\cite{yor}. This we defer to Section 3.3. 
Still, the problems will remain essentially the same ones in our present 
naive application. Thus  from (\re{3:9}), (\re{3:10}), (\re{3:11})  we obtain

\be
   E = \left. \frac{\delta K_0}{\delta N}\right |_r = \frac{\tilde E}{
    \sqrt{l}}  = \frac{\tilde E \sqrt{h}\,e^{-\alpha X}}{X^\prime}\;\; .
\la{3:15}
\ee
Eq. (\re{3:15}) refers to coordinate values $r$, the physical distance
$\lambda$ may be obtained by integrating 
\be
 d \lambda = \sqrt{h(r)} d r   \la{3:16}
\ee
wherever this is possible ($h > 0$). Thus (\re{3:15}) may also be written as

\be
E = \tilde E(\lambda)\; e^{-\alpha X} 
\left(\frac{dX}{d\lambda}\right)^{-1}\;\; . \la{3:17}
\ee
Neither $\tilde E(\lambda)$ nor the factor in (\re{3:17}) are
 gauge--independent. They still
contain  the arbitrary function $X(r(\lambda))$, even when we assume that a 
background with $C = 0$ is to be subtracted.  The situation is slightly
improved in the torsion--less case $\alpha = 0$, where (\re{3:8}) is to be
replaced by  
\ba
    \tilde E = 2\,(C - w) + v X + Y \frac{\partial w}{\partial Y}\nn
    w = \int\limits^{X(r)}_{X_0} v(y,Y) d y \;\; .\la{3:18}
\ea
Here the aforementioned subtraction of a background with $C = 0 $
produces a gauge--independent result $\tilde {E} = 2 C$ for any  $V =
v(X,Y)$. In any case, the problem of the gauge--dependent factor of $\tilde 
E$ in (\re{3:17}) remains.\\
Another way to see the same problem results from  eliminating $X$ by $f$ 
with (\re{2:25}) in (\re{3:17})
\be
E = \tilde E \frac{f^\prime}{\sqrt{h}} = \tilde E 
\frac{df}{d\lambda}\;\; ,
\la{3:19}
\ee
again measuring space by the physical distance $\lambda$. In the present 
coordinate system with a Killing field  (\re{2:21}) the factor of $\tilde 
E\,(\lambda)$ reflects the dependence of flat field on the space, in a way 
still to be determined by some {\sl additional} principle. We shall come 
back to that point in the next section.\\
As compared to the general torsion--less  
situation with (\re{3:18}), dilaton gravity in connection 
with a Noether current (\re{3:7}) needs a few comments, although the basic 
problems remain the same in that case. Here $Y = 0, v = v_0 = {\rm const.}$ in

(\re{3:18}), i.e. $\tilde E_{dil} = C - v_0\,X$. From the preceding discussion
for $\alpha \neq 0$ another modification
comes from the fact that the metric to be used in the ADM--setting is 
given by (\re{2:18}), i.e. contains another factor $(X/2)^{-1}$. Hence 
differentiation with respect to $N$ as in (\re{3:15}) or (\re{3:17}) provides a

factor $X/2\;$,  i.e. 

\be
E_{dil} = \frac{X}{X^\prime}\;\frac{\sqrt{h}}{2}\; (C - v_0\,X)\;\; .
\la{3:20}
\ee
Again a  subtraction of a background with $C = 0$ yields a  
gauge--independent $\tilde E_{dil}$. Now, for dilaton theory it is well--known

that the mass (prop. $C$) can be obtained by a appropriate limit relative to 
the dilaton vacuum with linearly rising (or decreasing) $\Phi$ at 
space--like distances \cite{bro}. In fact, with $h(\lambda) = 1$ in 
(\re{3:20}) 

\be
\lim_{\lambda \to \infty}\;\frac{d}{d\lambda}\; \ln X \propto \lim_{\lambda 
\to \infty}\; \frac{d\Phi}{d\lambda} \; = \; {\rm const.}
\la{3:21}
\ee
in an asymptotic sense determines (up to a constant factor) the mass. We 
shall take this as a hint for a possible extrapolation to the general case 
in Sect. 3.3. \\
The 'naive' Noether
current approach still did  require some asymptotic limit, but it, at least,
avoided the mass--shell difficulty in our first attempt (\re{3:7}). 
Therefore,  one may hope that a direct application of the very general
approach e.g.  of Wald \cite{wal} for Noether charges in covariant theories
may avoid those problems: The Lagrangian $\cal L$ in our action (\re{1:2})
is already a two--form as required in \cite{wal}.  Its variation by a
diffeomorphism with Lie--derivative $\Lambda_\xi$ produces a surface term

\ba
\delta {\cal L} &=& e.o.m. + d\Theta \nonumber\\
        \Theta &=& X^B\, \Lambda_\xi \, A_B \;\; ,
\la{3:22}
\ea
using the comprehensive notation for connections and target space 
coordinates of (\re{1:1}). Following the argument of \cite{wal}, 
$\Theta$ can be related to a Noether current one--form 

\be
j = \Theta - \xi \cdot {\cal L}
\la{3:23}
\ee
where the dot indicates  a contraction with  the first index of the 
two--form $\cal L$. It can be verified easily  from (\re{3:22}, 
\re{3:23}) with (\re{1:2}) that $j$ is exact on--shell:

\ba
j  &=& d Q + e.o.m. \nonumber\\
Q  &=& X^B\, \xi \cdot A_B
\la{3:24}
\ea
For $\xi$ we may choose the Killing vector $k$ of (\re{2:21}), i.e.  simply
selecting the $df$ components in the solution (\re{2:7}).  Within a patch
where $k$ is a time--like Killing vector, $Q$ may be evaluated at each
point of 'space' $X^3 = X$.  By comparing with the (variation of the)
symplectic one--form serving as a Hamiltonian \cite{wal} the (variation of
the) Wald's 'energy'--density on the (here zero--dimensional) boundary
becomes

\be
\delta E_{(W)} = \delta X^B\, k \cdot A_B = \delta X^B\, (A_B)_f \;\; .
\la{3:25}
\ee
However, we arrive at the desired gauge--independent result $E_{(W)} \propto 
C$ again only by not fully going on--shell: Note that the 
$f$--components of the solution (\re{2:7}) may be expressed formally as

\be
\left(A_B\right)_f \quad = \quad \frac{\partial\, C(X^A)}{\partial X^B}
\la{3:26}
\ee
{\sl as long as} $C$ in (\re{2:5}) is taken to be a function of the $X^A$ 
and not a constant. Then (\re{3:25}) may be integrated to yield 

\be
E_{(W)} \quad = \quad C \;\; ,
\la{3:27}
\ee
but we have the same mass--shell problem as with the approach leading 
to ${\cal T}$ above. The present argument following \cite{wal} also does 
not change in the case of dilaton gravity, the Noether charge being the 
constant $C$ and hence essentially coinciding with the black--hole mass
everywhere 
(and not only asymptotically). 

\subsection{Regge--Teitelboim Surface Term}

The problems encountered in the preceding Sections 3.1 and 3.2 may be 
summarized as follows: \\
Firstly, energy momentum in the Killing direction (\re{2:21}), and 
the closely related general Noether charge (\re{3:24}) in the sense of 
\cite{wal},  yield the gauge--independent conserved quantity $C$, however 
in each of these cases full use of the field equations (including $ C = 
{\rm const.}$) would not give any result. \\
By contrast the 'naive' Noether current (\re{3:7}) could be evaluated with the

full equations of motion yielding again a surface term. However,  that 
surface term showed already gauge--dependence which could be compensated 
by subtraction of a (nonflat) background (in the torsion--less case only). 
\\
Secondly, a problem 
arose if -- by analogy with a quasilocal energy \cite{wal} --- it was tried to

calculate the 'energy' referring to a proper ADM--setting from such a 
surface term. At best 'asymptotically' a relation to $C$ could be obtained 
in the special case of dilaton gravity. \\
What is usually called 'ADM--mass', especially within numerous applications 
in 2d theories, explicitly or implicitly is based upon the argument of Regge 
and Teitelboim \cite{reg} for a Hamiltonian formulation in a finite region of 
space: An appropriate surface term with (spacelike normal vector) must be 
added to the Hamiltonian so that the equations of motion are reproduced 
without the requirement of the vanishing variation on that surface. Clearly 
the choice of that surface determines the result in a profound way. Actually 
this argument can be considered to be somehow hidden in the first and third 
approaches of Sect.\ 3.1 and 3.2, as applied to our present d = 2 
theories. Here we use  it explicitly to isolate the origin of the mass--shell 
problem and to propose a simple remedy. In our 
case the Hamiltonian with momenta $ X^A = (X^-, X^+,X,Y)$ may be simply read 
off from the first order action (\re{1:2})

\be
  H = \int\limits^b_a dr[e_t^+ G^- +e_t^- G^+ +\omega_t G + A_t \tilde G +
  \partial_r O ]\;\; , \la{3:28}
\ee
fixing $t$ to be the Hamiltonian time. $H$ consists of the constraints
\ba 
     G^{\pm} &=& - \partial_{r} X^{\pm} \mp \omega_r X^{\pm} \pm e_r^{\pm}
V\nn
     G       &=& - \partial_{r} X + X^+ e_r^- - X^- e_r^+ \la{3:29}\\
     \tilde G &=& - \partial_r Y\nonumber
\ea 
and of a surface  density $O$, which should be related to the conserved 
quantity. With the abbreviations $\bar q_A = \{e_t^+, e_t^-, \omega_t, 
A_t\}$ the straightforward steps from (\re{1:2}) to (\re{3:28}) show that in
this case with 
\be
 O^{(0)} = \bar q_A X^A \la{3:30}
\ee
the validity of the Hamilton--Jacobi equations  with nonvanishing variations
on
the surface for the $X^A$ 
 from (\re{3:28}) are  verified easily. However, (\re{3:30}) exactly coincides
with the $K_0$ of (\re{3:7}) and (\re{3:7a}) in the preceding subsection which
has been found to be an unlikely candiate. \\
There are, however, other ways to implement this argument. One is the 
proposal  
to (linearly) recombine the constraints in (\re{3:28}) so that a surface
term appears naturally \cite{geg,fis}.  In our present completely general
case the proper combination of constraints can be deduced by a simple
algebraic argument.  The algebra of (first class) constraints $G^A$ and
$X^A$ closes \cite{gro}, and its center has two elements, $Q(X)$ which as a
function of $X^A$ coincides with $C$, but can be considered {\it not} to be
constant for the present, because the e.o.m.  are not fulfilled,  and
\be
  U(X)=(V G + X^+ G^- + X^-G^+)e^{\alpha X}\;\; .  \la{3:31}
\ee
The trivial further quantity $\tilde U = Y\tilde G$ involving the 
$U(1)$--field is not important here. Actually (\re{3:31}) generates the 
diffeomorphisms in 'space' \cite{fro} and may be expressed (still off--shell)
as  
\be
      U = - \partial_r\,Q(X) \;\; .   \la{3:32}
\ee
Following the steps of \cite{geg} in our completely general case (\re{3:28}),
we eliminate the Lorentz--constraint $G$ in favour of $C$ by (\re{3:31}) with
(\re{3:32}):
\be
    H = \int\limits^b_a d r\,\left[\omega_t \,(e_t^c - \frac{X^c}{V})G_c +
\frac{\omega_t
    e^{-\alpha X}}{ V} (-\partial_r Q)+\partial_r O^{(1)}\right] \la{3:33}
\ee
Now $ O^{(1)}$ is determined in such a way that with the Killing 
direction as 'time' the canonical equations of motion are 
reproduced for the finite end--points $a$ and $b$ of 'space' $r$. Varying 
the first terms in (\re{3:33}), only expressions involving $\partial X^3,
\partial X^a$ are critical. In order to compensate those by a total
derivative,  for the correct equations for $X$ an appropriate $ O^{(1)}$ must
be chosen
as 
\be
  O^{(1)}= \frac{\omega_te ^{-\alpha X}}{ V}\, Q(X)\;\; .   \la{3:34}
\ee
{\sl In the next step} we consider the on--shell limit $Q(X) = C = {\rm 
const.}$, 
expressing also the factor in front  of $Q$ by means of the solution for 
$\omega_t$ in (\re{2:7}):
\be
  O^{(1)} = \dot f\, C \la{3:35}
\ee
Actually, the same result obtains simply  requiring 
that in (\re{3:28}) the proper equations of motion  for the $X$ should
be reproduced --- without any recombination of constraints. From the 
Regge--Teitelboim argument then  the necessary condition for $O$ obviously is
\be
 \delta O = \bar q_A \delta X^A \la{3:36}
\ee
in the short--hand notation already used above. Relaxing again for the moment
the on--shell condition that $X^-$ is related to $X^+$ by (\re{2:5}) at fixed 
constant $C$, and treating $C = C(X^A)$ as a function, we again use the 
identity (\re{3:26}) which in our present notation reads
\be
    \bar q_A = \dot f\frac{\partial Q(X)}{\partial X^A}\; .\la{3:37}
\ee
Inserting (\re{3:37}) into (\re{3:36}) allows the immediate integration 
leading to (\re{3:35}) again. \\
Within this formulation, it is now easy to see how we are able to solve at 
least the problem of a gauge--independent {\sl surface} term: We just 
interpret that term containing (\re{3:37}) as given
'independently' from the full field equations (which would
require $C = {\rm const.}$).  We 'guess' it so as to yield the
proper Hamiltonian equations of motion for the $X^A$, because
that compensation works before the equations of motion are
obtained and before they are used to fix $C = {\rm const.}$\,. 
In retrospect we see that a similar argument should have been
introduced within the context of the approaches leading to
(\re{3:5}) and (\re{3:7a}), however it seems to be difficult to
find an equally obvious way to do this.\\ Eq.(\re{3:35})
immediately allows to make contact with the projections of
Noether charges according to (\re{3:2}) and (\re{3:4}) above. 
Fixing our coordinate system so that $df$ is 'time' ($dt$) in
(\re{3:35}) precisely corresponds to that projection producing a
result like (\re{3:27}).  \\ For the quasilocal energy the
problem, encountered in Sect.  3.2 remains, although now instead
of (\re{3:15}) and (\re{3:17}) in

\ba
E &=& \beta \, C\nonumber\\
\vert \beta \vert &=& \frac{e^{-\alpha X}}{(dX / d\lambda)} = 
\frac{e^{-\alpha X}\sqrt{h(r)}}{X^\prime}
\la{3:38}
\ea
the gauge--dependence reduces to the factor $\beta$, which persists even 
measuring length by the 
 physical scale $\lambda$. In 2d covariant models only in exceptional 
cases an asymptotically flat direction exists, e.g. in dilaton gravity where 
$\beta \to {\rm const.}$ asymptotically for the dilaton vacuum. However, 
already in that case a more careful definition would read for $r > r_c$ (or 
$\lambda > l_c$) 
\be
\vert \vert \beta \vert - {\rm c }\vert << 
 \vert p \vert^\alpha \; ,\la{3:39}
\ee
implying that above a certain space--like distance (again in a diagonal
metric $N_1 = 0$ for simplicity) the deviation from some constant $c$ is 
much smaller than the smallest parameter in the solution with a power $\alpha$

to match the dimension of the l.h.s. in (\re{3:39}). We observe that
(\re{3:39})
may be assumed to hold in a general 2d theory as well, but for a {\sl 
finite} interval (in 'physical space' $\lambda$) so that $r_0 < r < r_1$. 
The conditions to be fulfilled for such a region are most easily seen in 
Schwarzschild coordinates (\re{2:29}). From $\vert \beta(r) \vert \sim c$

\be
\vert\, c\,\vert \simeq \left\langle \frac{1}{\sqrt{l}} \right\rangle
\la{3:40}
\ee
follows  where the average value for the norm of the 
Killing vector in the interval $[r_0,r_1]$ appears, as indicated by the 
r.h.s.    Of course, for 
(generalized) dilaton theories $\vert l \vert$ has to be corrected by the 
proper field by definition. In both cases the 'average' metric becomes 
$g_{\alpha\beta} \sim \langle g_{\alpha\beta} \rangle = {\rm 
diag}\, (c^{-2}, - c^2)$, i.e. we have, as expected, flat space. \\
Obviously (\re{3:40}) only holds far from the zeros (horizons, 
singularities) of $\vert l \vert$ and may be well satisfied e.g. near 
(flat) extrema, saddle points etc.  Actually precisely those properties for 
a function which essentially coincides with $l\,(r) / X(r)$ in our present 
notation has been the basis of the general discussion of singularity 
structures in ref. \cite{klo} so that details for the existence of an
appropriate 
interval $[r_0, r_1]$ in a specific theory can be deduced easily for any 
model with any given set of parameters.

\section{Conclusion and Outlook}

Exploiting the properties of conserved charges in 2d covariant matterless 
theories in a comprehensive framework, we are able to critically analyze 
pseudo--tensor and 
ADM--methods to obtain quasilocal conserved quantities. In two dimensions 
the lack of asymptotic flatness is compensated by the fact that all 
(gauge--independent) conserved quantities are known, together with the 
general solution in arbitrary gauges. E.g. the mass of the black hole in 
dilaton gravity is just one particular very simple example for such a 
conserved quantity. In our present work we explicitly used for illustration 
an example where one conservation law $C_1 = const.$ refers to 'energy', and 
a second one is related to another U(1) gauge group ('field strength'). \\
Trying to determine Noether--type charge definitions and to apply the 
concept to quasilocal energy, we encountered two basic problems which are 
typical for matter-less d = 2 covariant theories in the generic case, 
including theories with dynamical torsion as well. One problem arose with 
the use of the full equations of motion, an additional one appeared together 
with a quasilocal energy definition. Solutions to both problems
could be obtained by
proper interpretation of the argument of Regge and Teitelboim for 
compensating terms on surfaces (in a finite distance) with a space--like 
normal  in the Hamiltonian approach although  that argument 
is not naively applicable in d = 2 theories: It is possible to 'guess' that 
surface term from the solution without actually going fully on--shell. This 
provided the solution for a (gauge--independent) Noether--charge, to be 
identified with $C$. By analogy with their asymptotic definition of 
quasilocal energy in dilaton gravity we propose a similar definition at 
finite distance, avoiding in this way the problem that, in general, the 
singularity structure has no 'flat' direction in 2d--theories. \\
 It is obvious from the present approach that no positivity for any such
'energy' can be expected.  From the mathematical point of view any sign of
$C_1$ is equally acceptable, albeit belonging to very different singularity
structures.  Of course, any statement about the (classical) stability of
such structures is possible only when additional interactions with
'genuine' matter are considered.  A single scalar field, allowing always the
reinterpretation as a dilaton is not enough.  This not only holds for the
original dilaton black hole \cite{bro,ban,man} but for any covariant 2d
theory.  In our class of models the second conserved quantity influences
the singularity structure at the same level as e.g.  the cosmological
constant.  It naturally appears as part of the 'energy' $C_1$.  The special
role of one single conserved quantity --- as opposed to the PSM-s
\cite{strb} --- clearly is typical for the peculiar choice of target space
appropriate for 2d covariant theories.  \\
In the course of our analysis we also encountered in one case (Sect.\ 3.1) 
the special role of the 'background' singularity structure in 2d covariant 
theories interacting with matter. Therefore it seems that precisely that 
structure assumes the place of an asymptotically flat 'background' in d = 4. 
If this were true in general, all 2d--modelling of GR involving 
time--dependent singularity structures in the interaction with matter etc. 
may suffer from a serious inherent defect. Fortunately known \cite{bro} and 
more recent \cite{klo} results at least for the dilaton black hole do not
support 
such a pessimistic view, at least in the case of theories of that type.\\
In the quantum case a genuine {\sl field} theory also only arises in 
interaction with matter. Without that only on a suitable compactified space 
isotopic to $S^1$ the finite number of zero modes precisely of the $C$-s covers

a quantum mechanical theory with a finite number of degrees of freedom. \\
Although $C_1$ (in our generic case) turns into a 'energy density', not 
necessarily constant in space and time anymore when matter is present 
\cite{kumd}, it retains its physical aspects related to the geometrical part of
the 
action --- very much like the mass parameter in the so far very most 
prominent case, the example of the dilaton black hole interacting with 
matter: E.g. in the energy momentum approach from eq.(\re{3:2}), 
generalizing even to an $\omega$--dependent $L^{(m)}$ with the r.h.s. of the 
second eq.(\re{2:2}), containing a one--form ${\cal S}^{(m)}$, the steps 
leading to (\re{3:3}) imply a relation 

\be
\tilde{\cal T} = {\cal T} + 2 V {\cal S}^{(m)} = dC
\la{3:41}
\ee
and thus again a generalization of the absolute conservation law for $C$ to 
a relation of type (\re{3:4a}).  $C$ now will vary in space and time, in
general.\\
We should emphasize again that the definition of $\cal T$ involves a 
matter--less Killing direction i.e. refers to a background which assumes the 
place of asymptotically flat space in customary applications to d = 4 GR. \\
We believe that our present work considerably broadens the range of possible 
starting points in that direction, at least by the almost limitless increase 
in conceivable singularity structures to be analyzed within this context.

\section*{Acknowledgement:}

The authors have profitted from discussions  with H. Balasin, T. Kl\"osch, P. 
Schaller and T. Strobl. This work has been supported by Fonds zur F\"orderung 
der wissenschaftlichen Forschung (Project P-10221--PHY). 



\begin{thebibliography}{99}
\bibitem[18]{int}	In the quantum case the $C_i$ are identified for nontrivial
topology on 
		the space $R \times S^1$  as the only genuine variables of the 
		theory [11,12]
\end{thebibliography}
\end{document}